\documentstyle[preprint,aps]{revtex}

\begin{document}
\title{\bf Symmetries and Universality Classes in 
Conservative Sandpile Models}
\author{Ofer Biham\footnote{E-mail: biham@flounder.fiz.huji.ac.il},
Erel Milshtein\footnote{E-mail: milstein@flounder.fiz.huji.ac.il}
and Sorin Solomon\footnote{E-mail: sorin@cc.huji.ac.il}}
\address{
Racah Institute of Physics, 
The Hebrew University, 
Jerusalem 91904, 
Israel}

\maketitle

\begin{abstract}
The symmetry properties which determine the critical exponents and universality
classes in conservative sandpile models are identified.
This is done by introducing a set of models, including all possible combinations 
of abelian vs. non-abelian, deterministic vs. stochastic and isotropic vs. 
anisotropic toppling rules.
The universality classes are determined by an extended set of critical exponents, scaling 
functions and geometrical features.
Two universality classes are clearly identified: (a) the universality class of abelian models 
and (b) the universality class of stochastic models. In addition, it is found that 
non-abelian models with deterministic toppling rules  exhibit non-universal behavior.
\end{abstract}


\newpage

Sandpile models were introduced about a decade ago as a paradigm of 
self organized criticality (SOC)~\cite{B87,B88,T88}. 
SOC provides a useful framework for the study of a large class of driven
non-equilibrium systems which dynamically evolve into a critical state.
At the critical state these systems exhibit long-range spatial and 
temporal correlations, which resemble the behavior at the equilibrium 
critical point. SOC was thus proposed as the mechanism underlying the
appearance of fractal structures and $1/f$ noise, which exhibit power-law
spatial and temporal correlations~\cite{B87,B88,T88}. 
The critical state of SOC systems  can be characterized by critical exponents,
 scaling functions and other geometric features.
To examine these properties a variety of sandpile models 
have been introduced
\cite{Z89,M91}
and their scaling properties
were studied both analytically 
\cite{V95,C97}
and numerically
\cite{L97}.
Numerical studies indicate the existence of a number of distinct universality 
classes of sandpile models~\cite{B96,M98}. However, a systematic 
classification relating the universality classes to the underlying symmetries 
has not been achieved.

In this letter we introduce a systematic framework for the classification of 
sandpile models into universality classes.
This framework is based on the fundamental symmetries which can be identified 
in sandpile models: (a) the abelian symmetry; (b) the rotational symmetry 
(isotropic, uniaxial, directed) and (c) deterministic vs. stochastic 
update rules.
Using this framework we identify the relevant parameters of sandpile models
and the assignment into universality classes.

We will first introduce the models.
Sandpile models are defined on a $d$-dimensional lattice of linear size $L$.
Each site {\bf i} is assigned a dynamic variable $E({\bf i})$ which 
represents some physical quantity such as energy,  stress, etc. A  configuration
$\left \{E({\bf i}) \right \}$ is called {\it stable} if
for all sites $E({\bf i}) < E_c$,  where $E_c$ is a threshold value.
The evolution between stable configurations is by the following 
rules:
 (i) Adding energy.
Given a stable configuration $\{E({\bf j})\}$ we select a 
site ${\bf i}$ at random and increase $E({\bf i})$ by some amount 
$\delta E$.
When an unstable configuration is reached rule (ii) is applied.
(ii) Relaxation rule.
If $E({\bf i}) \geq$ 
$E_c$,   
relaxation takes place and energy is distributed in the following way:

\begin{eqnarray}
\label{def}
E({\bf i}) & \rightarrow & E({\bf i}) - \sum_{\bf e}\Delta 
E({\bf e})  \nonumber \\				
E({\bf i}+{\bf e}) & \rightarrow & E({\bf i}+{\bf e})+
\Delta E({\bf e}), 
\end{eqnarray}
where ${\bf e}$ are a set of (unit) vectors from the site ${\bf i}$
to some neighbors. 
As a result of the relaxation, $E({\bf i}+{\bf e})$ for one or more of
 the neighbors may exceed the threshold $E_c$.  
The relaxation rule is then applied until a stable 
configuration is reached.
The sequence of relaxations is an avalanche which propagates 
through the lattice.

Since the parameters $\delta E$ and $E_c$  are irrelevant to the scaling 
behavior~\cite{D90,D92,D94}, the critical exponents depend only on the
vector $\Delta E$,  to be termed {\it relaxation vector}. 
For a square lattice with relaxation to nearest neighbors (NN) it is of 
the form $\Delta E=(E_N, E_E, E_S, E_W)$,  where $E_N, E_E,  E_S$ and $E_W$ are
the amounts transferred to the northern,  eastern,  southern and western NN's respectively. 
In the Bak-Tang-Wiesenfeld (BTW) model, $E_c=4$, $\delta E=1$ and $\Delta E=(1, 1, 1, 1)$. If an active 
site with  $E({\bf i}) > E_c$ is toppled, it would not become empty after 
the topple had occurred. 
In the Zhang model~\cite{Z89}, for which $E_c=1$ and $0<\delta E<1$, the relaxation vector is given by $(b, b, b, b)$, 
where $b=E({\bf i})/4$
 and $E({\bf i})$ 
 is the amount of energy in the active site before
the topple had occurred. Obviously,  the site ${\bf i}$
remains empty after toppling.
In the random relaxation models introduced by Manna~\cite{M91} a set of neighbors is randomly chosen 
for relaxation.
Such models are specified by a set of relaxation vectors,  each 
vector being assigned a probability for it application.
For example,  a possible realization of a two-state Manna model includes six relaxation 
vectors (1, 1, 0, 0), (1, 0, 1, 0), (1, 0, 0, 1), (0, 1, 1, 0), (0, 1, 0, 1) and (0, 0, 1, 1),  
each one applied with a 
probability of $1/6$. 
A {\it time step} (of unit {\it time}) is defined as the relaxation of all the
sites having  $E({\bf i}) \geq  E_c$,  after the completion of the previous time step. 

To identify the relevant parameters of sandpile models we will consider the 
symmetries of the relaxation rules.
{\it The abelian symmetry:} A model is said to be {\it abelian} if the 
configuration after an avalanche, is independent of the 
 order in which the relaxation of the active sites was performed.
The BTW model was shown to be abelian~\cite{D90}.
The Manna models~\cite{M91} are not abelian because they contain a 
random choice of the toppling direction. As a result,  they develop different scenes of toppling 
that depend on the order of relaxation of the active sites even within a single time step. The Zhang
model is also non-abelian. This can be seen,  when two active NN 
 sites are toppled within the same time step: the site that was toppled last remains empty while 
the other one is non-empty. This shows that the final configuration depends on the order.
{\it The rotational symmetry:}
A model is said to be {\it isotropic } if in each topple 
the energy is divided equally between the NN's.
 The BTW and the Zhang models are isotropic by definition while the Manna models
 are non-isotropic.
{\it The deterministic vs. stochastic feature:}
 A model is said to be {\it deterministic} 
if the toppling rule is deterministic, otherwise it is stochastic.
The BTW and the Zhang models are deterministic 
 while the Manna models are stochastic.

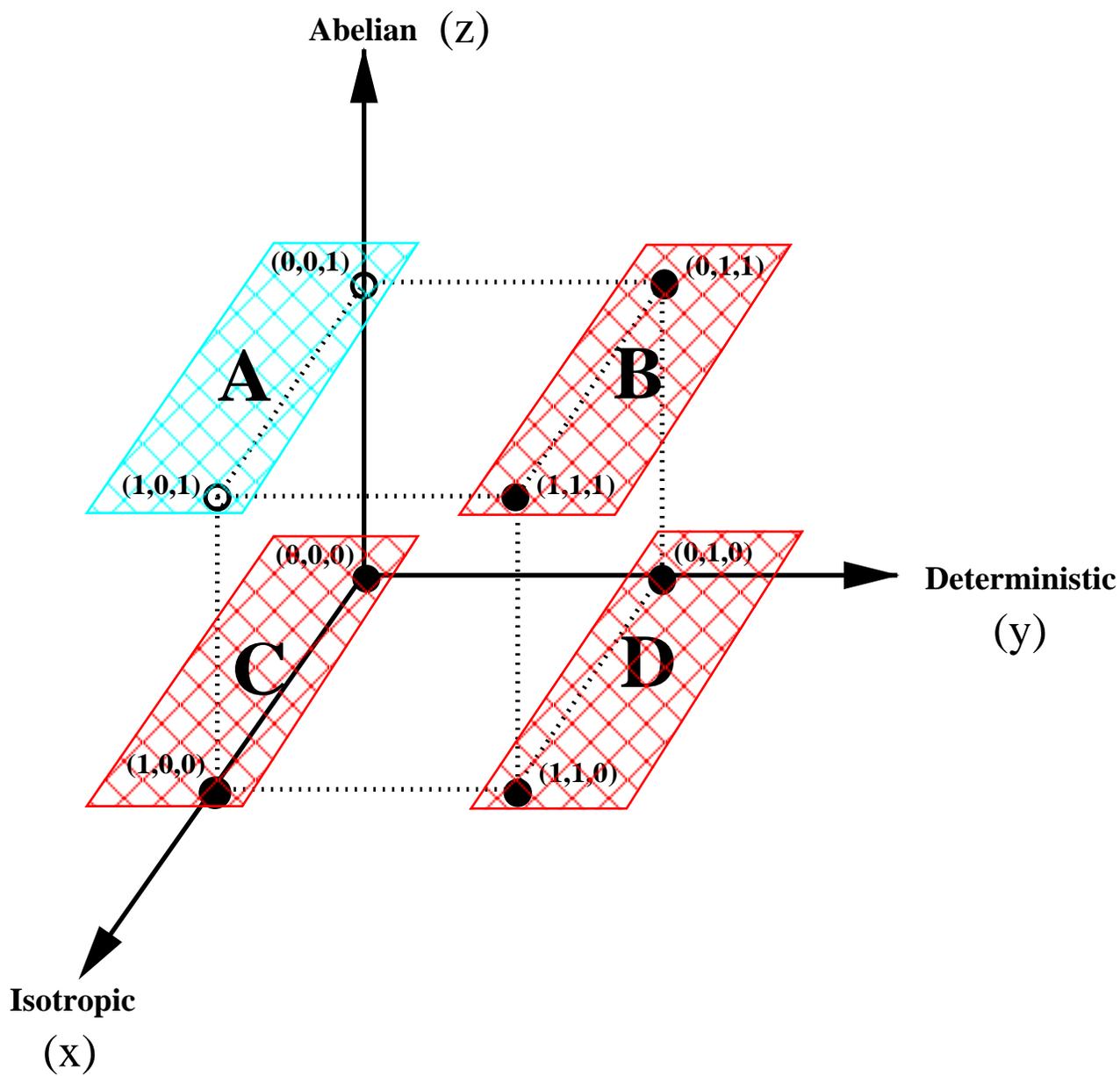
\begin{figure}
\caption{Classification diagram of sandpile models into the corners of the unit
three dimensional cube $(x,y,z)$, where $x=1$ ($0$) for isotropic (anisotropic),
 $y=1$ ($0$) for deterministic (stochastic) and  $z=1$ ($0$) for abelian
(non-abelian) models. Four classes are identified: 
A. An empty class of implausible models;
B. The abelian universality class (BTW);
C. The stochastic universality class (Manna);
D. A class of non-universal models (Zhang).}
\label{phaze_uni}
\end{figure}

The three properties described above form eight possible combinations which can be 
graphically represented on the corners of a three dimensional cube (Fig.~\ref{phaze_uni}).
The $x$ axis represents the rotational symmetry, where $x$=$1$ ($x$=$0$)
 for isotropic (anisotropic) models.
The $y$ axis represents the deterministic vs. stochastic feature where
  $y$=$1$ ($y$=$0$) for 
deterministic (stochastic) models.
The $z$ axis represents the abelian symmetry, where $z$=$1$ ($z$=$0$) for 
 abelian (non-abelian) models.
Each corner of the cube is specified by its coordinates $(x,y,z)$.
The BTW model belongs to the $(1,1,1)$ corner, the Zhang  model belongs to the
 $(1,1,0)$ corner and  
the Manna models belong to the $(0,0,0)$ corner. 
To systematically identify the relevant parameters we need to introduce models that belong 
to each of the eight corners and examine their scaling properties.
We first note that an abelian model cannot be stochastic and vice versa.
This is due to the fact that in a stochastic model (for a given seed of the random number
generator) the actual moves to be 
performed in a given time step depend on the order in which they are performed.
As a result, stochastic models cannot be abelian and  the 
corners $(0,0,1)$ and $(1,0,1)$  in Fig.~\ref{phaze_uni} remain vacant.

We will now introduce a new set of models which will later be used for a 
systematic study
of the effects of the three properties described above on the critical behavior.
These models fall into two groups: variations of the Zhang 
model and variations of the  Manna model.
We will first introduce variations of the Zhang model.

\noindent
{\it Generalized  Zhang} (GZ) : in this model 
the Zhang relaxation vector is modified into $\Delta E= (b,b,b,b)$
 where $b=pE({\bf i})/4$ and 
$0 < p \leq 1 $ is a pre-determined constant, 
such that only a fraction $p$ of the energy in site ${\bf i}$ is 
distributed to its neighbors~\cite{modification}. Since none of the properties is 
changed this model remains in the $(1,1,0)$ corner in Fig.~\ref{phaze_uni}.
In the limit $p \rightarrow 0$ the model 
becomes abelian as the order of relaxations becomes irrelevant. In this limit it is 
termed the {\it Abelian Zhang} (AZ) model, which shifts into
 the $(1,1,1)$ corner in Fig.~\ref{phaze_uni}.
 This resembles the situation in rotations of a rigid body in the three 
dimensional space, where infinitesimal rotations are abelian 
while finite rotations are not~\cite{G80}. Moreover, within this analogy, the BTW  model 
may correspond to the group of rotations by $180^{\circ}$ around the $x$, $y$ and $z$ axes,
which is also abelian.

\noindent
{\it Parallel-Update Zhang} (PZ) : this is a variation of the Zhang model, in which 
 all the topplings in a given time step are performed  {\it simultaneously}, 
namely each unstable site ${\bf {i}}$ distributes
 $E({\bf i})/4$ to each nearest neighbor,
where  $E({\bf i})$ is its energy after the previous time step was completed.
The model can thus be considered as abelian and assigned to the $(1,1,1)$ corner.

\noindent
{\it Stochastic Zhang} (SZ) :
in this model the Zhang relaxation vector is modified into 
$\Delta E= (b,b,b,b)$ where $b=rE({\bf i})/4$ and in each toppling event 
$r$ is chosen randomly in the range $0 < r \leq 1 $.
Therefore, the SZ model belongs to the 
$(1,0,0)$ corner.

\noindent
{\it Alternating  Uni-Axial Zhang} (AUZ): this model has two relaxation vectors: $(c,0,c,0)$ 
used in odd time steps and $(0,c,0,c)$ used in even time steps, where $c=E({\bf i})/2$. 
The basic features of the Zhang model are maintained except for the isotropy, and 
therefore it shifts to the  $(0,1,0)$ corner.

\noindent
We will now introduce models which are variations of the Manna models.

\noindent
{\it Uni-Axial Manna} (UM): this is a restricted version of
the Manna two-state model.
It  includes two equally probable relaxation vectors:
$(1, 0, 1, 0)$ and $(0, 1, 0, 1)$
which are both uniaxial. The UM model has the same three properties as the ordinary 
Manna models, and it thus belongs to the $(0,0,0)$ corner. It only differs  from
the Manna models in the fact that is does not include directed moves.

\noindent
{\it Alternating  Uni-Axial Manna} (AUM): this
model has the same relaxation
vectors as the UM model, but it is not stochastic, since the $(1,0,1,0)$ vector is used for
all topplings in odd time steps and the $(0,1,0,1)$ vector is used for even time steps.
This deterministic rule that decouples the horizontal and vertical directions makes the model 
abelian, and it thus belongs to the 
 $(0,1,1)$ corner.

To obtain a complete characterization of the models introduced above we have performed extensive 
computer simulations of all the models and calculated an extended
set of characterization measures.
These measures include the distribution exponents, the geometric exponents, as well as 
scaling functions and geometric features of the avalanche~\cite{M98}.
The {\it distribution exponents} $\tau_x$ characterize the distribution of various avalanche parameters.
It is found that $P(x) \sim x^{-1-\tau_x}$, where $x$ may represent the avalanche size (s), area (a)
or time (t). The {\it geometric exponents} $\gamma_{xy}$ relate the distribution of these quantities,
and are defined in terms of the conditional expectation values
 $E[x|y] \sim y^{\gamma_{xy}} $
where $x,y \in \{ s,a,t \}$
\cite{C91,C93}. 
The scaling functions describe the time evolution of the avalanche size $S(t)$ 
and area growth rate $A(t)$
during the avalanche, averaged over a large number of avalanches.
According to the dynamic scaling assumption, each one of these functions can be
written in the general scaling form :
\begin{equation}
X(t) = K_X\langle t \rangle_X^{-\alpha_X} f_X(\mu)
\end{equation} 
where $\mu = t/ \langle t \rangle_X$, $X \in \{ S, A\}$ and 
\begin{equation}
\langle t \rangle_X=\frac{\sum_{t} tX(t)}{\sum_{t} X(t)}.
\end{equation} 

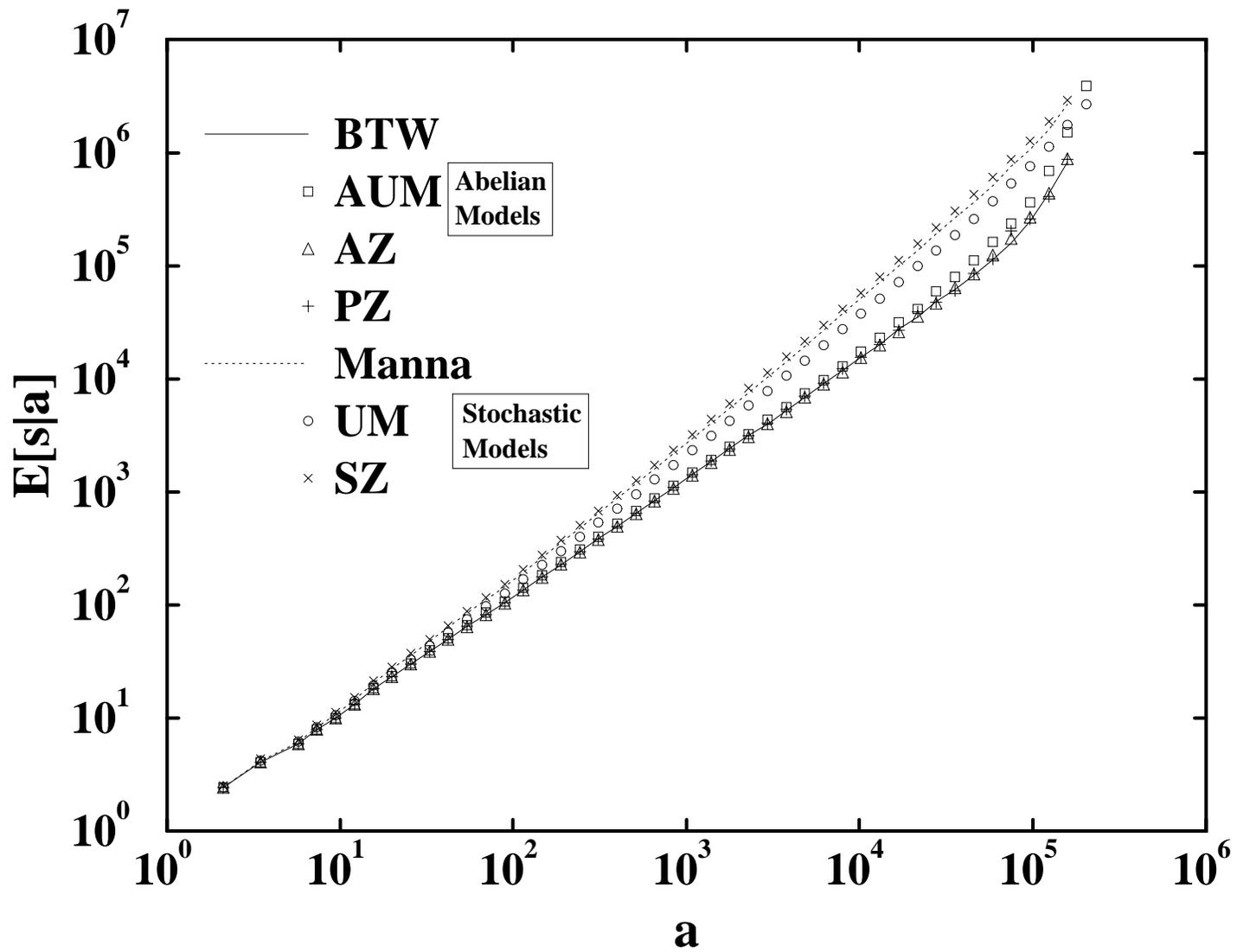
\begin{figure}
\caption{The conditional expectation values $E[s|a]$ vs. $a$ which yields the geometrical critical
exponents $\gamma_{sa}$ for the set of abelian and stochastic models. For the abelian models we find
that $\gamma_{sa} = 1.05 \pm 0.01$, while for the stochastic models 
$\gamma_{sa} = 1.24 \pm 0.02$. The small parameter $p$ in the 
AZ model is $p=0.005$.}
\label{gamma_all}
\end{figure}

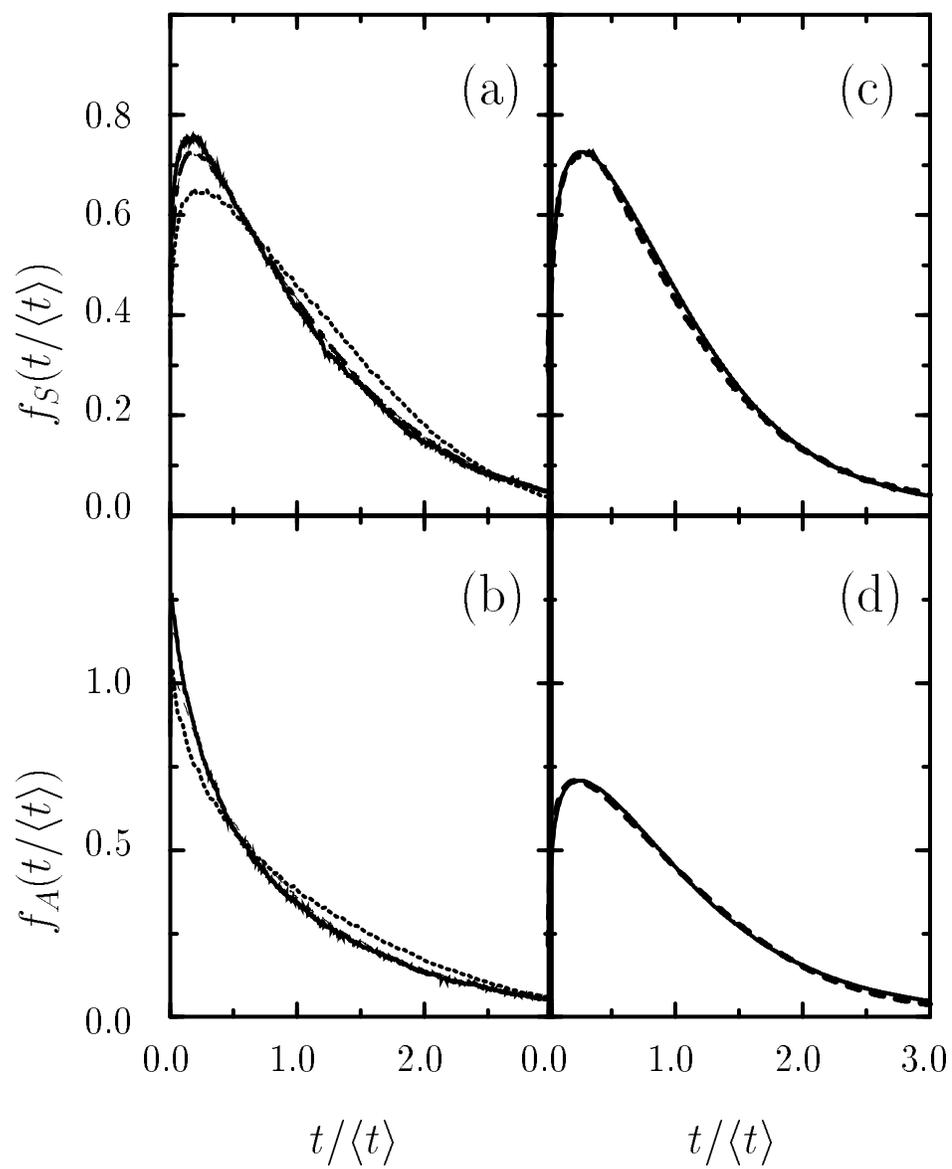
\begin{figure}
\caption{The scaling functions for the abelian and stochastic models.
 (a) $f_S(t/\langle t \rangle)$ for the stochastic models, 
Manna (---), SZ (- -) and UM ($\cdots$);
  (b) $f_A(t/\langle t \rangle)$  for the same set of stochastic models. While the scaling functions for the first 
two models coincide, the third one has some deviation;
 (c) $f_S(t/\langle t \rangle)$ for the abelian models, 
BTW (---) and AZ (- -);
 (d) $f_A(t/\langle t \rangle)$ for the same set of abelian models. The scaling functions for 
these two models perfectly coincide.}
\label{scaling}
\end{figure}

The conditional expectation values  $E[s|a]$ vs. $a$ are shown in Fig.~\ref{gamma_all} for the 
abelian models of class $B$ (see Fig.~\ref{phaze_uni}) and the stochastic 
models of class $C$. We find that the exponent $\gamma_{sa}$ for the abelian 
models, namely the Alternating  Uni-Axial Manna (AUM), 
Abelian Zhang (AZ) and Parallel-Update Zhang (PZ) models
coincides with its value for the BTW model, $\gamma_{sa} = 1.05 \pm 0.01$.
For the stochastic models, namely the  Uni-Axial Manna (UM) and 
Stochastic Zhang (SZ) models,  $\gamma_{sa}$
coincides with the Manna  value $\gamma_{sa} = 1.24 \pm 0.02$.
The fact that the BTW  and Manna models belong to different universality classes was pointed 
out before~\cite{B96}. However, the results presented here indicate a considerable degree of 
universality within each class and also attribute it to the abelian symmetry in class $B$ and 
to the stochastic dynamics in class $C$.
The scaling functions for the abelian and stochastic models are shown in 
Fig.~\ref{scaling}. In the stochastic class we observe very good coincidence between the Manna 
and Stochastic Zhang (SZ)   models both for  $f_S(\mu)$ and $f_A(\mu)$. The Uniaxial Manna (UM) 
model somewhat deviates from the other two, although it exhibits the same qualitative shape.
In the abelian class we observe perfect coincidence between the BTW  and the Abelian Zhang (AZ) 
for both  $f_S(\mu)$ and $f_A(\mu)$. These results provide further evidence for Universality 
within each of the two classes.

To further characterize the avalanche structure we examined the function 
$f({\bf i})$, that provides the number of toppling events at site ${\bf i}$ 
during the avalanche~\cite{M98}.
For the abelian models,  we observe a shell structure in which all sites which relaxed at least $n+1$ 
times form a connected cluster with no holes which is contained in
 the cluster of sites which relaxed at least $n$ times.
The Stochastic models exhibit a random avalanche structure 
with many peaks and holes~\cite{B96}.

As we pointed out before, class $A$ is empty since a stochastic model cannot 
be abelian. We will
now consider the remaining class, $D$. It turns out that systems in class $D$ exhibit non-universal
critical behavior. 
The exponents $\gamma_{sa}$ and $\tau_s$
of the generalized Zhang (GZ) model 
vary continuously as the parameter $p$ is lowered
from the Zhang values 
$\gamma_{sa}= 1.6 \pm 0.05 $
and
$\tau_s = 1.25 \pm 0.01 $
at $p=1$
to the BTW values
$\gamma_{sa}= 1.05 \pm 0.01$
and
$\tau_s = 1.09 \pm 0.01$
at $p \rightarrow 0$.
For the models of class $D$, the 
avalanche structures are intermediate between 
those of the abelian and stochastic models and
depend on the model parameters, such as the parameter $p$ in the GZ model.
For example, as $p \rightarrow 0$
the avalanche structure converges to the ordered shell structure of the 
BTW model.
Moreover, for this class the average size and area growth functions during the avalanche do not
collapse into scaling functions. This indicates that models in class $D$ do not exhibit all
the features of critical behavior which appear in classes $B$ and $C$.

In summary, we have performed a systematic study of critical behavior, relevant parameters and 
universality classes in sandpile models which exhibit self organized criticality. We introduced an 
extended set of models, including all possible combinations of abelian vs. non-abelian,
deterministic vs. stochastic and isotropic vs. anisotropic toppling rules. To characterize the 
critical behavior we have used an extended set of critical exponents, particularly relying on the
geometric exponents, which were found to be most useful, in addition to scaling functions and 
geometric features of the avalanche. Two universality classes were clearly identified and attributed 
to the underlying symmetry properties: the universality class of abelian models (which includes
the BTW model) and the universality class of stochastic models (which includes the Manna models).
In addition it was found that the class of deterministic models which are 
non-abelian (which includes the Zhang model)
exhibits non-universal behavior.

A number of promising theoretical frameworks, 
based on the fixed scale transformation approach
\cite{V95}
and on the dynamic renormalization group approach
\cite{C97}
have been introduced in recent years for the study
of universality in sandpile models.
We believe that extending these approaches to include
the relevant symmetry properties examined here would
greatly improve our theoretical understanding of SOC.

We thank M. Paczuski for useful discussions and correspondence.

\newpage

\newpage


\begin{references}
\bibitem{modification}
		Note that such modification of the relaxation rule applied to the BTW or 
		Manna models would simply  amount to a shift and rescaling of the 
		energy unit.
\end{references}
\end{document}